\documentclass[12pt]{article}
\usepackage{amsmath, amsfonts, subfigure}
\usepackage{graphicx, hyperref}
\PassOptionsToPackage{pagebackref}{hyperref}
\usepackage{natbib}
\usepackage{url} % not crucial - just used below for the URL 
\usepackage{xcolor}

\newcommand{\blind}{0}

\renewcommand{\(}{\left(}
\renewcommand{\)}{\right)}

\newcommand{\DKL}{D_{\mathrm{KL}}}
\newcommand{\argmax}{\operatornamewithlimits{argmax}}

\newcommand{\EE}{\mathbb{E}}

\begin{document}

\def\spacingset#1{\renewcommand{\baselinestretch}%
{#1}\small\normalsize} \spacingset{1}

%%%%%%%%%%%%%%%%%%%%%%%%%%%%%%%%%%%%%%%%%%%%%%%%%%%%%%%%%%%%%%%%%%%%%%%%%%%%%%

\if0\blind
{
  \title{\bf A Likelihood-Free Approach to Goal-Oriented Bayesian Optimal Experimental Design}  
\author{Atlanta Chakraborty \thanks{atlantac@umich.edu
    }\hspace{.2cm}\\
    Mechanical Engineering, University of Michigan, Ann Arbor, MI 48109\\
    Xun Huan \thanks{xhuan@umich.edu} \\
     Mechanical Engineering, University of Michigan, Ann Arbor, MI 48109\\
    and \\
    Tommie Catanach \thanks{tacatan@sandia.gov} \\
    Sandia National Laboratories, Livermore, CA 94550}
    \date{}
  \maketitle
} \fi

\if1\blind
{
  \bigskip
  \bigskip
  \bigskip
  \begin{center}
    {\LARGE\bf Title}
\end{center}
  \medskip
} \fi

\bigskip
\begin{abstract}
Conventional Bayesian optimal experimental design seeks to maximize the expected information gain (EIG) on model parameters. However, the end goal of the experiment often is not to learn the model parameters, but to predict downstream quantities of interest (QoIs) that depend on the learned parameters. And designs that offer high EIG for parameters may not translate to high EIG for QoIs. 
\emph{Goal-oriented} optimal experimental design (GO-OED) thus directly targets to maximize the EIG of QoIs. 

We introduce LF-GO-OED (likelihood-free goal-oriented optimal experimental design), a computational method for conducting GO-OED with nonlinear observation and prediction models. 
LF-GO-OED is specifically designed to accommodate implicit models, where the likelihood is intractable. In particular, it builds a density ratio estimator from samples generated from approximate Bayesian computation (ABC), thereby sidestepping the need for likelihood evaluations or density estimations. 
The overall method is validated on benchmark problems with existing methods, and demonstrated on scientific applications of epidemiology and neural science.  

%\textbf{160/200 words} 
\end{abstract}

\noindent%
\vfill

\newpage
\spacingset{1.75} % DON'T change the spacing!

\section{Introduction}
\label{sec:intro}

Optimal experimental design (OED) \citep{Pukelsheim2006,Atkinson2007,Ryan2016,Rainforth2023,Huan2024}
has found widespread interest and usage across engineering, science, and medicine, where a carefully designed experiment can often bring substantial benefits.

A Bayesian approach to OED, emphasized in the influential papers by~\citet{Lindley1956} and \citet{Chaloner1995} among others, further incorporates the notion of Bayesian uncertainty and update by adopting a design criterion based on the mutual information, or expected information gain (EIG), of model parameters.

Let $\Theta \in \boldsymbol{\Theta} \subseteq \mathbb{R}^{n_{\Theta}}$ denote the random vector of unknown model parameters, $Y \in \mathcal{Y} \subseteq \mathbb{R}^{n_{Y}}$ the observation data from an experiment, and $\xi\in \Xi \subseteq \mathbb{R}^{n_{\xi}}$ the experimental design variable (e.g., the conditions under which the experiment is conducted). When an experiment is conducted under design $\xi$ yielding observations\footnote{We use upper case to denote random variable/vector, and lower case to denote realized value.} $Y=y$, the prior density of $\Theta$ is updated to the posterior density via Bayes rule:
\begin{align}
p(\theta|y,\xi) = \frac{p(y|\theta,\xi)\, p(\theta)}{p(y|\xi)},
\label{e:Bayes}
\end{align}
where $p(\theta)$ is the prior density of $\Theta$, $p(\theta|y,\xi)$ is the posterior density, $p(y|\theta,\xi)$ is the likelihood, and $p(y|\xi)$ is the marginal likelihood. 
The OED design criterion---EIG on $\Theta$~\citep{Lindley1956}---can be written as
\begin{align}
    U_{\Theta}(\xi) = \iint p(y,\theta|\xi) \log  \frac{p(\theta|y,\xi)}{p(\theta)}  \,\text{d}\theta \,\text{d}y = \EE_{Y|\xi} \left [\DKL(p_{\Theta|Y,\xi}\,||\,p_{\Theta}) \right]. \label{eq:objective}
\end{align}  
Notably, such a criterion can also be interpreted from a decision-theoretic point of view as an expectation of a utility function; in the case of \eqref{eq:objective}, the utility is the Kullback--Leibler (KL) divergence $\DKL$ from the prior to the posterior, quantifying the entropy reduction on $\Theta$ from observing $y$ at design $\xi$. Hence, $U_{\Theta}$ is also called the \emph{expected utility}, with the subscript $\Theta$ here indicating it is based on the EIG of $\Theta$. The OED problem then entails finding the design that maximizes the expected utility: $\max_{\xi\in \Xi} U_{\Theta}(\xi)$.

In many situations, however, learning $\Theta$ is not the ultimate goal of the experiment. Instead, one may be interested in other quantities of interest (QoIs) that depend on $\Theta$ through either a deterministic prediction model (i.e., parameter-to-QoI mapping) $Z=H(\Theta)$ or a stochastic one $Z=H(\Theta,\mathcal{E})$, with $\mathcal{E}$ being a stochastic variable associated with the QoI predictions. 
One such example is when training data-driven models such as deep neural networks, where the model parameters are abstract and not of direct interest, while its predictions are more interpretable and often of ultimate interest. 
In these cases, EIG on the QoIs $Z$ would be more appropriate to serve as the design criterion~\citep{Bernardo1979}:
\begin{align}
    U_Z(\xi) 
=\iint p(z,y|\xi) \log  \frac{p(z|y,\xi)}{p(z)}  \,\text{d}y \,\text{d}z =\EE_{Y|\xi}[\DKL(p_{Z|Y,\xi}\,||\,p_Z)], 
 \label{eq:goalobjective}
\end{align}
where $p(z)$ and $p(z|y,\xi)$ are the prior-predictive and posterior-predictive densities, respectively.
\citet[Theorem 1]{Bernardo1979} shows that $U_Z(\xi) \leq U_{\Theta}(\xi)$, with equality holding when $H$ is bijective.
The (predictive) \emph{goal-oriented OED (GO-OED)} problem then entails finding the design that maximizes this new expected utility, $\max_{\xi\in \Xi} U_{Z}(\xi)$.

We focus on GO-OED in this paper. 
While the EIG formulation for $Z$ is straightforward to state, its computation is non-trivial. In particular, estimating and optimizing $U_Z$ requires evaluating the 
posterior-predictive density $p(z|y,\xi)$. For example, one possible numerical approach~\citep{zhong+sch22} involves sampling the posterior via Markov chain Monte Carlo (MCMC)~\citep{Robert2004,Andrieu2003}, 
computing the predictive QoIs for each sample through $H$, and estimating the posterior-predictive density using a density estimation technique---a costly procedure overall.
Furthermore, with implicit models, the likelihood (i.e., the density $p(y|\theta,\xi)$) is intractable to evaluate, and even methods such as MCMC can no longer be used. Intractable likelihood has motivated the development of simulation-based inference, such as approximate Bayesian computation (ABC)~\citep{sisson+fb18}, that solely requires the ability to draw samples from $p(y|\theta,\xi)$ without needing density evaluations. Similarly, these principles may be imported to GO-OED.

In this paper, we introduce LF-GO-OED (likelihood-free goal-oriented optimal experimental design), a likelihood-free computational method for GO-OED, for estimating $U_Z(\xi)$. LF-GO-OED (a) leverages ABC for sample-generation to sidestep the need for likelihood evaluation, and (b) directly estimates the \emph{density ratio} in \eqref{eq:goalobjective} to avoid any density estimation. Our key contributions can be summarized as follows.
\begin{enumerate}
\item We present LF-GO-OED, a new computational method for performing GO-OED with nonlinear observation and prediction models.

\item We enable LF-GO-OED to be suitable for handling implicit or intractable likelihood.

\item We validate LF-GO-OED on benchmarks with other established algorithms, and demonstrate its effectiveness on scientific applications in epidemiology and neural sciences. 
\end{enumerate}

The remainder of this paper is structured as follows. Section~\ref{sec:Literature} first provides a literature review pertaining to GO-OED, likelihood-free approaches and their uses in OED contexts. Section~\ref{sec:Methodology} then details the numerical method of LF-GO-OED, which includes ABC sampling and training of density ratio estimators. Section~\ref{sec:Examples} provides results demonstrating LF-GO-OED on a number of benchmarks and application problems. Finally, the paper ends with concluding remarks in Section~\ref{sec:Conclusion}. Additional details supporting the numerical examples can also be found in the Appendix.

\section{Related Literature}\label{sec:Literature}

The formulation of GO-OED 
is rooted in classical L- and D$_\text{A}$-optimal designs for linear models, which seek to optimize the trace and log-determinant, respectively, of the Fisher information matrix under a linear combination of the model parameters~\citep{Atkinson2007}. 
I-, V-, and G-optimal designs further expand these criteria to linear regression models~\citep{Atkinson2007}. In the Bayesian setting, \citet{Attia2018} present a gradient-based approach to finding D$_\text{A}$- and L-optimal designs where $Y$ and $Z$ both linearly depend on $\Theta$ and where all distributions remain Gaussian. 
\citet{wu2021efficient} advance scalable offline-online decompositions and low-rank approximations to reduce the complexity for high-dimensional QoIs, also with linear models. 
\citet{Butler2020} further propose OED4P, an algorithm that estimates an objective based on the predictive-pushforward distribution, but does not fall under a Bayesian framework. 
More recently, \cite{zhong+sch22} propose a method for nonlinear Bayesian GO-OED by using MCMC and kernel density estimation (KDE) to compute the posterior-predictive density; however this approach tends to be  computationally expensive and difficult to scale to high dimensions. 
Importantly, all of these methods require an explicit likelihood, and cannot handle intractable models. 

For OED with implicit likelihoods under a non-goal-oriented setting, several papers utilize ABC~\citep{sisson+fb18}, for instance, in \citet{drovandi+p13, hainy+mw14, price+brt16}. In particular, ABC obtains approximate posterior samples by employing a distance measure for comparing summary statistics between simulated and observed data, thereby only requiring samples from the likelihood model without any likelihood evaluations. 
However, \cite{dehideniya+dm18} show that ABC is appropriate primarily for lower-dimensional settings.
Alternatives to ABC
have also been explored
broadly under simulation based inference methods \citep{cranmer+bl20}, including 
Bayesian synthetic likelihood \citep{price+dln18} and MCMC variants without likelihoods \citep{marjoram+mpt03, sisson+ft07}. Their usage for posterior inference is typically expensive, and applying them for OED would become even more costly 
due to the need for repeated calculations across different $y$ and $\xi$. Faster alternatives based on neural networks have been recently introduced in \citet{papamakrios+sm19, bremer+lpc20, durkan+pm18,zammitmangion+sh24}, but have yet been tested within OED; we refer interested readers to a \href{https://github.com/smsharma/awesome-neural-sbi}{repository} detailing these likelihood-free methods and their applications.

A different strategy to tackle OED with implicit models is to  directly approximate the density ratio in \eqref{eq:objective} and bypass the posterior density entirely~\citep{kleinegesse+g19}. For example, likelihood-free inference by ratio estimation (LFIRE) \citep{owen+dckg22}  estimates the density ratio in \eqref{eq:objective}  by training a logistic regression classifier to identify whether a sample $y$ is generated from the likelihood $p(y|\theta,\xi)$ or from the marginal likelihood $p(y|\xi)$. While~\cite{kleinegesse+g19} build a classifier for each individual $\theta$, it is possible to amortize a single classifier over $\Theta$ \citep{cranmer+pl16}. 
Such amortized ratio estimators have been used in other applications,
for instance, to evaluate the acceptance ratios in MCMC~\citep{hermans+bl20}. In another work by \cite{kleinegesse+gutmann20}, they consider using a neural network to maximize a lower bound on the mutual information. Numerous other variational bound approaches have also been investigated in the context of OED, for instance by \cite{foster+jbhtrg19, pacheco+f19}. Recently, \cite{dahlke+zp23} propose using moment matching projection methods instead of gradient-based optimization to expedite computations for optimal designs under exponential family assumptions. The iDAD algorithm by \citet{Ivanova2021} and vsOED by \cite{Shen2023a} further accommodate implicit models in a sequential OED setting. Elsewhere, \cite{ryan+dp16, overstall+m20} suggest building an auxiliary likelihood model (also called indirect inference) for Bayesian OED for implicit models. 

In the context of GO-OED with implicit models,  
\cite{kleinegesse+g21} build on their earlier work of neural network-based mutual information lower bound
\citep{kleinegesse+gutmann20} and
generalize the design criterion to the mutual information between an arbitrary quantity of interest and the data, making it suitable for implicit models with goals of model discrimination and predictions. One such lower bound that relates to LFIRE is the Jensen--Shannon divergence (JSD), where it is shown that maximizing the JSD is equivalent to minimizing the objective in LFIRE. Similarly, one could build an amortized neural network to estimate the density ratio in \eqref{eq:goalobjective}. However, we find that a very large neural network is required to accommodate the amortized data, and meticulous tuning of the hyper-parameters to be difficult and time-consuming; therefore we do not pursue this path in this paper.

\section{Methodology}\label{sec:Methodology}
We begin by noting that the GO-OED expected utility from \eqref{eq:goalobjective} can be expressed in two equivalent forms:
\begin{align}
  U_Z(\xi) 
&= \iint p(z,y|\xi) \log  \frac{p(z|y,\xi)}{p(z)}  \,\text{d}y \,\text{d}z 
 \label{oneform}\\
 &= \iint p(z,y|\xi) \log  \frac{p(y|z,\xi)}{p(y|\xi)}  \,\text{d}y \,\text{d}z.  \label{secondform}
\end{align}
A straightforward approach is to create nested Monte Carlo (MC) estimators for them:
\begin{align}
U_Z(\xi)  &= \iint p(z,y|\xi) \log  \frac{\int p(z|\theta)p(\theta|y,\xi) \,\text{d}\theta}{\int p(z|\theta')p(\theta') \,\text{d}\theta'}  \,\text{d}y \, \,\text{d}z \approx \frac{1}{N}\sum_{i=1}^N \log \frac{\frac{1}{J}\sum_{j_1=1}^Jp(z_i|\theta_{j_1}^{(2)})
 }{\frac{1}{K}\sum_{k=1}^Kp(z_i|\theta_k^{(4)}) 
 }, \label{mc1} \\
 U_Z(\xi)  &=\iint p(z,y|\xi) \log  \frac{\int p(y|\theta,\xi)p(\theta|z) \,\text{d}\theta}{\int p(y|\theta',\xi)p(\theta') \,\text{d}\theta'}  \,\text{d}y \, \,\text{d}z \approx  \frac{1}{N}\sum_{i=1}^N \log \frac{\frac{1}{J}\sum_{j_2=1}^Jp(y_i|\theta_{j_2}^{(3)},\xi)
 }  {\frac{1}{K}\sum_{k=1}^Kp(y_i|\theta_k^{(4)},\xi) 
 },
 \label{mc2}
\end{align}
where for both forms, the samples $(z_i,y_i)\sim p(z,y|\xi)$ are generated by first sampling $\theta_i^{(1)}\sim p(\theta)$ and then drawing $y_i \sim p(y|\theta_i^{(1)},\xi)$ and $z_i \sim p(z|\theta_i^{(1)})$; 
in the log-numerators, $\theta_{j_1}^{(2)} \sim p(\theta|y_i,\xi)$ and $\theta_{j_2}^{(3)} \sim p(\theta|z_i)$; and in the log-denominators, $\theta_{k}^{(4)} \sim p(\theta)$. However, direct evaluations of the nested MC estimators would be expensive and more crucially, require ability to evaluate the likelihood, $p(y|\theta,\xi).$ 

Instead, our strategy is to construct direct estimators for the density ratios:
\begin{align}
 U_Z(\xi) =\iint p(z,y|\xi) \log  \frac{p(z|y,\xi)}{p(z)}  \,\text{d}y \, \,\text{d}z
 \approx \frac{1}{N}\sum_{i=1}^N \log \hat{r}_1(z_i,y_i) =: \widehat{U}_{Z,1}(\xi), \label{dr1} \\
 U_Z(\xi) =\iint p(z,y|\xi) \log  \frac{p(y|z,\xi)}{p(y|\xi)}  \,\text{d}y \, \,\text{d}z \approx \frac{1}{N}\sum_{i=1}^N \log \hat{r}_2(z_i,y_i) =: \widehat{U}_{Z,2}(\xi),
 \label{dr2}
\end{align}
where again $(z_i,y_i)\sim p(z,y|\xi)$. The ratio estimators $\hat{r}_1$ and $\hat{r}_2$ are built in a supervised learning manner. Training $\hat{r}_1$ will require generating samples of $Z$ from its posterior-predictive $p(z|y,\xi)$ and prior-predictive $p(z)$; similarly, training $\hat{r}_2$ will require generating samples of $Y$ from its posterior-predictive $p(y|z,\xi)$ and prior-predictive $p(y|\xi)$. 
Typically $Z$ is lower dimensional and without additionally taking into account of the underlying geometry of $Z$, $Y$, and $\Theta$, \eqref{dr1} may be easier to work with when $\dim(Y)\ll \dim(Z)$ since the dimension being conditioned would be lower,  
while \eqref{dr2} may be preferable when $\dim(Y)\gg \dim(Z)$ following the same argument. Working with posteriors conditional on lower-dimensional data has been shown to require less computational resources and time than that conditional on higher-dimensional data \citep{vono+pd22}. We will detail the sampling methods and construction procedures for density ratio estimators $\hat{r}_1$ and $\hat{r}_2$ in the next subsection.

Once the density ratio estimators become available, whether $\hat{r}_1$ or $\hat{r}_2$, an approximate optimal design can be found by solving the optimization problem
\begin{align}
    \xi^{\ast} \in \underset{\xi \in \Xi}{\argmax} \  \widehat{U}_{Z,1}(\xi) \quad \text{or} \quad \xi^{\ast} \in \underset{\xi \in \Xi}{\argmax} \ \widehat{U}_{Z,2}(\xi),\label{optimal design}
\end{align}
for example via grid search in $\Xi$ when it is low-dimensional (e.g., $\dim(\xi) \leq 3$) or via other 
optimization algorithms for higher-dimensional design spaces.

\subsection{Posterior Sampling Using ABC}\label{sec:ABC}

In order to train the density ratio estimators $\hat{r}_1$ and $\hat{r}_2$, training samples need to be drawn from $p(z|y,\xi)$ and $p(y|z,\xi)$, respectively. Sampling mechanisms for these distributions are not straightforward, and in fact each encompasses a posterior sampling problem. Sampling $z_j\sim p(z|y,\xi)$ requires first drawing a sample $\theta_j\sim p(\theta|y,\xi)$ and then sample $z_j\sim p(z|\theta_j)$. Similarly, sampling $y_j\sim p(y|z,\xi)$ requires first drawing a sample $\theta_j\sim p(\theta|z)$ and then sample $y_j\sim p(y|\theta_j,\xi)$. 
Ordinarily, sampling $p(\theta|y,\xi)$ and $p(\theta|z)$ can be performed using MCMC, but MCMC requires likelihood evaluations. We thus adopt likelihood-free inference methods of ABC~\citep{sisson+fb18} for this sampling task.
 
ABC methods use simulations in place of likelihood evaluations. For example, in order to sample from a posterior $\theta_j\sim p(\theta|y_{\text{obs}},\xi)$, candidate samples are first generated from the prior distribution $\theta' \sim p(\theta)$ and $y' \sim p(y|\theta',\xi)$. Then, the data sample $y'$ is reduced to low-dimensional summary statistics $S(y')$ and compared with observed summary $S(y_{\text{obs}})$. This step offers an advantage in handling high-dimensional data by reducing it to lower-dimensional summary statistics. Only candidate samples falling within a specified threshold of a distance metric, i.e., $D(S(y'), S(y_{\text{obs}}))<\epsilon$, are retained as the approximate posterior samples. 
We utilize the Euclidean metric in our illustrations and adopt a neural network adjustment ABC  \citep{blum+f10}. This adjustment involves employing a neural network as a nonlinear regression method to correct for the samples that do not produce an acceptable match.
We perform leave-one-out cross-validation on samples to determine the threshold associated with the lowest prediction error for each sample and tolerance value, using the function \texttt{cv4abc} in the \texttt{abc} package in R \citep{abc}. The appendix provides further implementation details on the choice of the threshold for all numerical examples in our paper.

\subsection{Density Ratio Estimators}\label{sec:DR}

Once we have the ability to generate samples from $p(z|y,\xi)$ or $p(y|z,\xi)$ using the techniques in Section~\ref{sec:ABC}, the goal now is to construct the density ratio estimator from these samples. Consider a general density ratio $r(x) = \frac{p(x)}{q(x)}$ with $q(x)>0$, $\forall x\in \mathcal{X}$
where $\mathcal{X}$ is the set of all possible realizations of $X$,
and where we have independent and identically distributed (i.i.d.) samples $x^p_i \sim p(x)$ and $x^q_i \sim q(x)$ for $i=1,\dots,N$.
We seek to train $\hat{r}(x)\approx r(x)$ using the data set $\{x^p_i,x^q_i\}_i$.
Ultimately, we adopt the uLSIF (unconstrained least-squares importance fitting) density ratio estimator implemented in the \texttt{densratio} package \citep{densratio} in R; below we introduce uLSIF in the context of other density ratio estimator choices, along with its advantages.

One possible approach is to train a probabilistic classifier, such as the logistic regression classifier in LFIRE, as the density ratio estimator. Probabilistic classifiers have been shown
to be optimal among a class of semi-parametric estimators in terms of minimizing the asymptotic variance of their maximum likelihood estimator \citep{sugiyama+sk10}; however, this holds only when the model is correctly specified, which is often not the case in practice.
When analyzing the estimator performance in terms of
$\EE_{\{x^p_i,x^q_i\}}[\DKL(p_X\,||\,\hat{r}_{\{x^p_i,x^q_i\}}q_X)]$---the expectation is taken over the training samples $\{x^p_i,x^q_i\}_i$,
and the subscript in $\hat{r}_{\{x^p_i,x^q_i\}}$ reminds the estimator's dependence on training data---\citet{kanamori+ss10} prove that when both $p(x)$ and $q(x)$ belong to the exponential family, estimators derived from probabilistic classifiers perform better than those from directly modeling $r(x)$ and minimizing the above KL divergence. However, the applicability of this assumption is generally difficult to verify in practice;
when violated, the KL divergence based estimators are, by construction, better performing than those from the probabilistic classifier approach.

Regarding estimator parameterization, \citet{sugiyama+nkbk07} propose KLIEP (Kullback--Leibler importance estimation procedure) that expresses the ratio estimator as a weighted sum of basis functions: $\hat{r}(x)= \sum_{m=1}^{M} \beta_m \phi_m(x)$, where the optimal values of $\beta=(\beta_1, \cdots, \beta_M)$ are determined by minimizing the KL divergence loss:
\begin{align}
  \min_{\beta \in \mathbb{R}^M} 
   \DKL (p_X\,||\, \hat{r}\, q_X)] &= \min_{\beta \in \mathbb{R}^M} \int p(x) \log \frac{p(x)}{\hat{r}(x) q(x)} \,\text{d}x\label{kliep}\\
   &= \min_{\beta \in \mathbb{R}^M}  \left[\int p(x) \log \frac{p(x)}{q(x)} \,\text{d}x - \int p(x) \log \hat{r}(x) \,\text{d}x \right]. \label{e:beta_KL2}
\end{align}
Since the first term in \eqref{e:beta_KL2} does not depend on $\beta$, the minimization statement in its MC form with samples $x_i \sim p(x)$ becomes
\begin{align}
    \max_{\beta \in \mathbb{R}^M} \frac{1}{N} \sum_{i=1}^N \log \hat{r}(x_i) =  \max_{\beta \in \mathbb{R}^M} \frac{1}{N} \sum_{i=1}^N \log \left(\sum_{m=1}^M \beta_m \phi_m(x_i) \right) \label{kliepopti}.
\end{align} 
A popular choice for the basis functions is Gaussian kernels
\begin{align}
\phi_m(x)= K(x, x_m)= \exp\( - \frac{|| x-x_m||^2}{2 \sigma^2}\), 
\end{align}
where $x_m$ includes points drawn from $p(x)$ or $q(x)$ or both,
and $\sigma$ is a tunable kernel width hyperparameter. 
Under this basis function choice, additional constraints are imposed: $\beta_1, \dots, \beta_M>0$ from non-negativity of the density ratio, and $\frac{1}{N} \sum_{j=1}^N \sum_{m=1}^M \beta_m \phi(x_j')=1$ to ensure that $\hat{r}(x)q(x)$ is a proper probability density, where $x_j' \sim q(x)$. 
The choice of basis functions has been observed to significantly impact the performance of KLIEP \citep{sugiyama+nkbk07}, and other basis settings are also worth exploring.

\cite{kanamori+hs09} later show that the KL minimization is inefficient due to the nonlinear logarithmic function induced by the divergence measure. As an alternative, they reformulated \eqref{kliepopti} as a convex quadratic optimization problem, resulting in a regularized least squares fitting approach known as the uLSIF. 
Correspondingly, the loss in \eqref{kliep} becomes the squared loss between the true density ratio and the estimated density ratio:
\begin{align}
&\min_{\beta \in \mathbb{R}^M} \frac{1}{2}\int \left(\frac{p(x)}{q(x)}- \hat{r}(x) \right)^2 q(x) \,\text{d}x 
\\
=& \min_{\beta \in \mathbb{R}^M} \frac{1}{2}\int \left(\frac{p(x)}{q(x)} \right)^2 q(x) \,\text{d}x -\int p(x) \hat{r}(x) \,\text{d}x+ \frac{1}{2}\int \hat{r}(x)^2 q(x) \,\text{d}x. \label{ulsif}
\end{align}
Similarly dropping the first term that does not depend on $\beta$, the minimization statement in its MC form can be succinctly written as 
\begin{align}
    \min_{\beta \in \mathbb{R}^M} \frac{1}{2} \beta^{\top} \hat{H} \beta -\hat{h}^{\top}\beta+\frac{\lambda}{2} \beta^{\top} \beta, \label{ulsifopti} 
\end{align}
where $\hat{H} \in \mathbb{R}^{M\times M}$ is a matrix whose entries are $\hat{H}_{ij}= \frac{1}{M} \sum_{m=1}^M K(x_j', x_m)K(x_j', x_i)$ with $x_j' \sim q(x)$ and $x_i, x_m \sim p(x); \hat{h} \in \mathbb{R}^{M}$ is a vector with entries $\hat{h}_i= \frac{1}{M}\sum_{m=1}^M K(x_i,x_m )$; and $\frac{\lambda}{2} \beta^{\top} \beta $ is a regularization term. The regression problem in \eqref{ulsifopti} has known stable analytical solution $\hat{\beta}=(\hat{H}+\lambda I_M)^{-1}\hat{h}$. 
Hyperparameters $\sigma$ and $\lambda$ can be selected through cross-validation. 

Building upon the uLSIF, \cite{liu+ycs13} introduce RuLSIF (Relative uLSIF) that is more robust in situations when $q(x)$ approaches zero 
(and so $\hat{r}(x)= \frac{p(x)}{q(x)}$
approaches infinity). This estimator is then replaced by the $\beta$ relative density ratio estimator, $\hat{r}_{\beta}(x)= \frac{p(x)}{\beta p(x)+ (1-\beta) q(x)}.$ When $\beta=0$, $\hat{r}_0(x)=\hat{r}(x)$; as $\beta \to 1$, the ratio estimator becomes smoother and is bounded by $1/\beta$ even when $p(x)/q(x)$ is unbounded. For all of our examples in this paper, since our prior is well-defined and appears in the denominator, $\beta=0$ suffices and reverts RuLSIF back to uLSIF. 
Therefore, for the numerical studies in this paper, we adopt the uLSIF density ratio estimator implemented in the \texttt{densratio} package in R \citep{densratio}.

\section{Numerical Examples}
\label{sec:Examples}

We illustrate the new LF-GO-OED method through a number of numerical examples. 
The first is a one-dimensional (1D) benchmark problem where the QoI is set equal to the parameter. Consequently, the expected utility values, and optimal design, from LF-GO-OED should coincide with those from non-goal-oriented OED which we obtain using nested MC~\citep{Ryan2003}. The purpose of this first example is thus to demonstrate the validity of LF-GO-OED. 
The second is a two-dimensional (2D) benchmark problem with a non-trivial QoI, where we compare LF-GO-OED to the MCMC-based GO-OED approach from \cite{zhong+sch22}. These benchmarks are followed by deploying LF-GO-OED on two scientific applications involving implicit models: 
the stochastic SIR model from epidemiology
and the stochastic FitzHugh--Nagumo (FHN) model
used for neural science. 

\subsection{Nonlinear 1D Benchmark}
\label{ss:nonlinear1D}

Consider the nonlinear observation model from \cite{huan+m11}:
\begin{eqnarray}
    Y=\Theta^3\xi^2+ \Theta \exp\left(-|0.2-\xi|\right) + \mathcal{E}, \label{toy1d}
\end{eqnarray}
where $\mathcal{E} \sim \mathcal{N}(0,10^{-4})$, prior $\Theta\sim \mathcal{U}(0,1)$, and $\xi \in[0,1]$. 
The tractable likelihood, i.e., $p(y|\theta,\xi)=p_{\mathcal{E}}(y-[\theta^3\xi^2+\theta\exp\left(-|0.2-\xi|\)])$, 
permits the non-goal-oriented expected utility $U_{\Theta}$ from \eqref{eq:objective} to be estimated via an established nested MC estimator \citep{Ryan2003}.
For LF-GO-OED, we set the QoI to equal to the model parameter, i.e., $Z=\Theta$ and therefore $U_Z(\xi)=U_{\Theta}(\xi)$, and estimate $U_Z$ using the estimator in \eqref{dr2}, $\widehat{U}_{Z,2}$, trained from 1000 samples.

\begin{figure}
\centering
\includegraphics[width=0.7\linewidth]{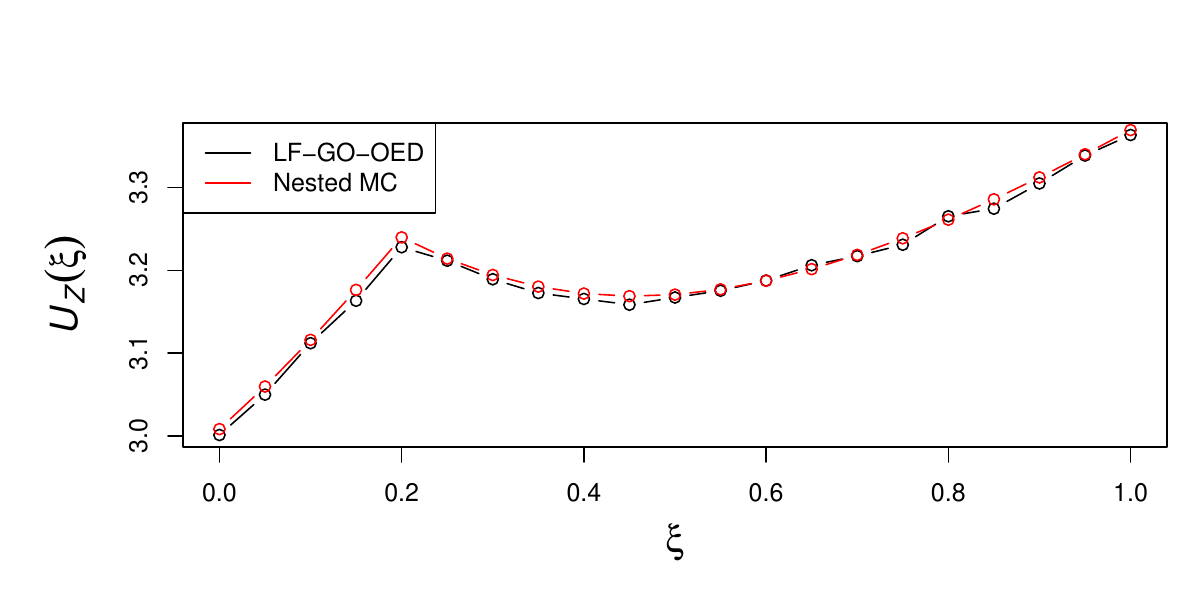} 
\caption{Nonlinear 1D benchmark. $U_{\Theta}$ estimated from nested Monte Carlo (red) and $U_Z$ (with $Z=\Theta$) estimated from LF-GO-OED estimator $\widehat{U}_{Z,2}$ (black). }
\label{fig:toy1d}
\end{figure}

Figure~\ref{fig:toy1d} shows the expected utility estimates using the two methods evaluated at 21 equally spaced design points in $\xi \in[0,1]$. LF-GO-OED is able to produce expected utility estimates that are near-identical to those from nested MC, validating the correctness of the algorithm in the non-GO-OED setting. We observe the least informative design to be at $\xi=0$ and two local maxima are at $\xi=0.2$ and $\xi=1$. These can be intuitively explained from the form of \eqref{toy1d}: at $\xi=0$ the observation variable $Y$ is dominated by the noise $\mathcal{E}$ and hence not useful for inferring $\Theta$, while at $\xi=0.2$ and $\xi=1$ the first and second terms are respectively maximized and thus offer the greatest signal-to-noise ratios. 

\subsection{Nonlinear 2D Benchmark}
\label{ss:2D}

Next consider a 2D benchmark from \citet{zhong+sch22} with a nonlinear model:
\begin{align}
    \begin{bmatrix} Y_1 \\ Y_2\end{bmatrix} &= \begin{bmatrix}\Theta_1^3\xi_1^2+ \Theta_2 \exp\(-|0.2-\xi_2|\) + \mathcal{E}_1 \\ 
     \Theta_2^3\xi_1^2+ \Theta_1 \exp\(-|0.2-\xi_2|\) + \mathcal{E}_2\end{bmatrix}, \label{eq:toy2d}
\end{align} 
where $(\mathcal{E}_1,\mathcal{E}_2) \sim \mathcal{N}(0,10^{-4}\mathbb{I}_2)$, prior $(\Theta_1,\Theta_2)\sim \mathcal{U}([0,1]^2)$, and $(\xi_1,\xi_2)\in[0,1]^2$.
The QoI for LF-GO-OED is set to the Rosenbrock function: $Z=(1-\Theta_1)^2+5(\Theta_2-\Theta_1^2)^2$.  We estimate $U_Z$ using the estimator in \eqref{dr1}, $\widehat{U}_{Z,1}$, to avoid needing to use a 1D $Z$ to infer a 2D $\Theta$ when sampling $p(\theta|z)$. $\widehat{U}_{Z,1}$ is trained using 1000 samples.
The tractable likelihood again enables us to validate LF-GO-OED against an existing GO-OED method based on MCMC and KDE \citep{zhong+sch22} that requires likelihood evaluations.  

\begin{figure}[htbp]
{\subfigure[LF-GO-OED]{\includegraphics[width=0.5\linewidth]{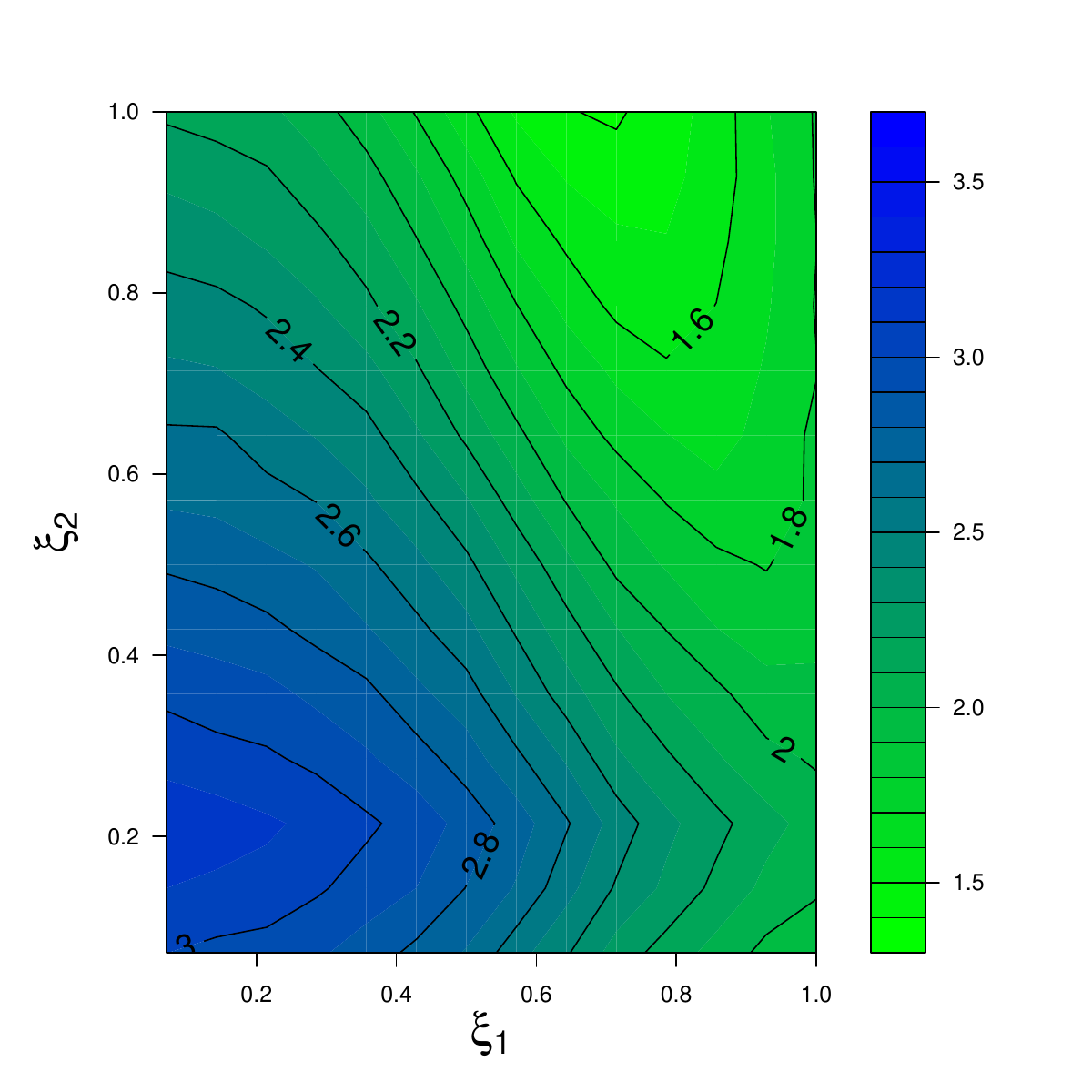}
      \label{fig:drtoy2d}%
    \qquad
    }
    }
 \subfigure[MCMC-based GO-OED]{\includegraphics[width=0.5\linewidth]{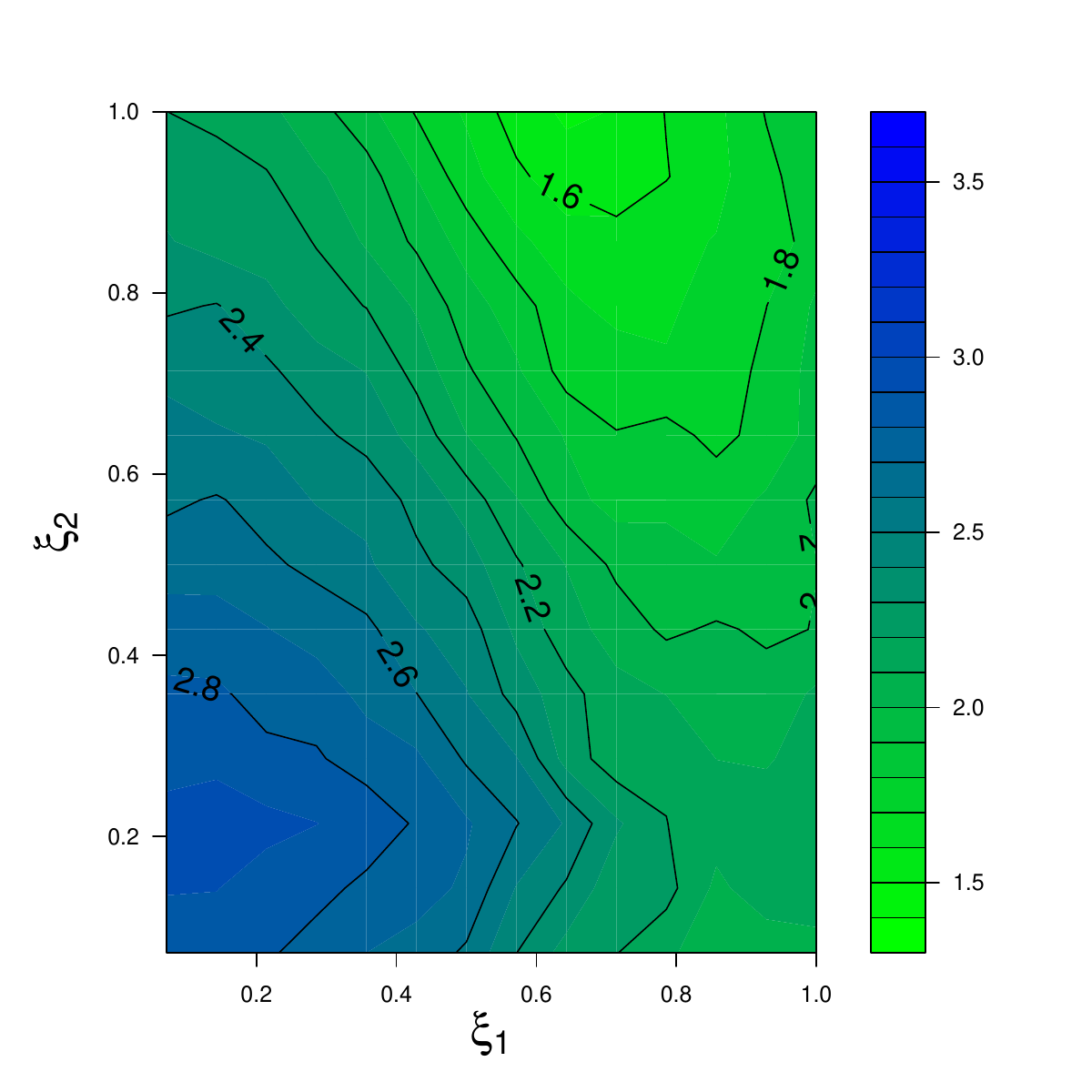}
\label{fig:dnmctoy2d}%}
}
 \label{fig:toy2d}
 \caption{Nonlinear 2D benchmark. $U_Z$ contours where the QoI is the Rosenbrock function, estimated from LF-GO-OED estimator $\widehat{U}_{Z,1}$ (left) and MCMC-based GO-OED (right).}
\end{figure}

Figure~\ref{fig:toy2d} presents the expected utility contours plotted on a $21\times 21$ uniform grid in the 2D design space. 
The optimal goal-oriented design for the Rosenbrock QoI found by LF-GO-OED is around $(\xi_1^{\ast},\xi_2^{\ast})=(0,0.2)$ with a corresponding $\widehat{U}_{Z,1}$ of 3.17. In comparison, the MCMC-based GO-OED \citep{zhong+sch22} arrives at an optimal design also at $(\xi_1,\xi_2)=(0,0.2)$ with estimated $U_Z$ of 2.97. The lower $U_Z$ estimate from the latter method likely stems from the diffusive nature of its KDE steps. Nonetheless, results from both approaches agree very well. This example thus provides validation of LF-GO-OED in the goal-oriented setting. 

In the appendix, we provide additional information about the computational cost of LF-GO-OED while varying parameter dimension and number of samples used in evaluating $\widehat{U}_{Z,1}$.

\subsection{Stochastic SIR Model for Epidemiology}

The SIR (Susceptible-Infectious-Recovered) model is a popular mathematical model for simulating and understanding the dynamics of infectious diseases. The stochastic variant of SIR, which is an intractable model, has been previously studied in the context of non-goal-oriented OED by \cite{kleinegesse+g19,kleinegesse+g21}. 
In this model, at time $t$ there is a population of $S(t)$ individuals who are in the Susceptible state, $I(t)$ individuals in the Infectious state, and $R(t)$ individuals in the Recovered state, totalling $N$ individuals. 
These population states evolve according to
\begin{align}
S(t+\Delta t) &= S(t)-\Delta S, \hspace{5.5em} \text{with } \Delta S \sim \text{Bin}(S(t), \beta I(t)/N),\\
I(t+\Delta t)&= I(t)+\Delta S-\Delta I, \hspace{3em}\text{with } \Delta I \sim\text{Bin}(I(t), \gamma),\\
R(t+\Delta t)&= R(t)+ \Delta I,
\end{align}
where $\Delta t$ is the discretized time step, and the changes $\Delta S$ and $\Delta I$ are stochastic following binomial distributions, respectively parameterized by the probability a susceptible individual becomes infectious, $\beta I(t)/N$, and the probability an infected individual recovers, $\gamma$.
The two global model parameters are therefore $\beta$ and $\gamma$. 
The likelihood is intractable in this case due to the 
nonlinear transmission dynamics and the 
stochastic evolution of the states. 
This model thus serves well to demonstrate the applicability of LF-GO-OED that can bypass the need for any likelihood evaluation or density estimation.

In our implementation, we employ $\Delta t= 0.01$, initial condition $(S(0), I(0), R(0))= (490,10,0)$, and endow the parameters with priors $\beta \sim \mathcal{U}(0,0.5)$ and $\gamma \sim \mathcal{U}(0,0.5)$. 
The design problem entails selecting a single time $\xi=t$ at which observation data $S(\xi)$, $I(\xi)$, and $R(\xi)$ can be accessed. 

\begin{figure}[htbp]
  \label{fig:sir}
  
  {\subfigure[$Z=\Theta$; optimal at $\xi^{\ast}=0.4, \widehat{U}_{Z,1}(\xi^{\ast})=2.07$ ]{\label{fig:sirparaminfnew}%
\includegraphics[width=0.44\linewidth]{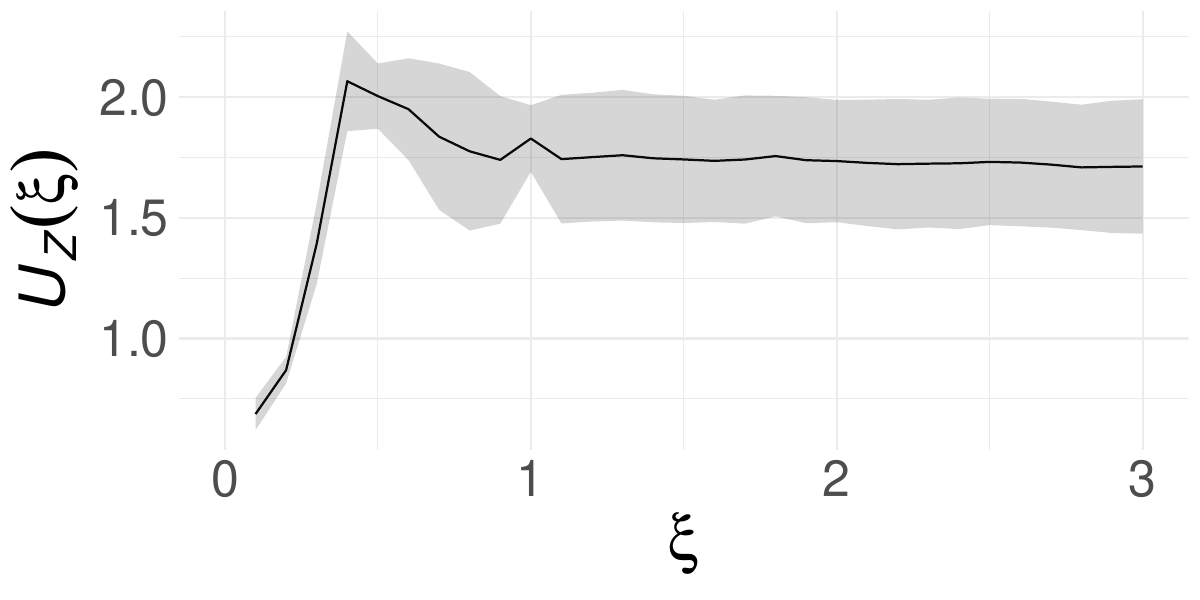}%
    \qquad
    }
    \qquad
    \subfigure[$Z= R(0.3)+R(0.4)+R(0.5)$; optimal at $\xi^{\ast}=0.2, \widehat{U}_{Z,1}(\xi^{\ast})=0.64$]{\label{fig:sirgoalrecov}
    
    \includegraphics[width=0.44\linewidth]{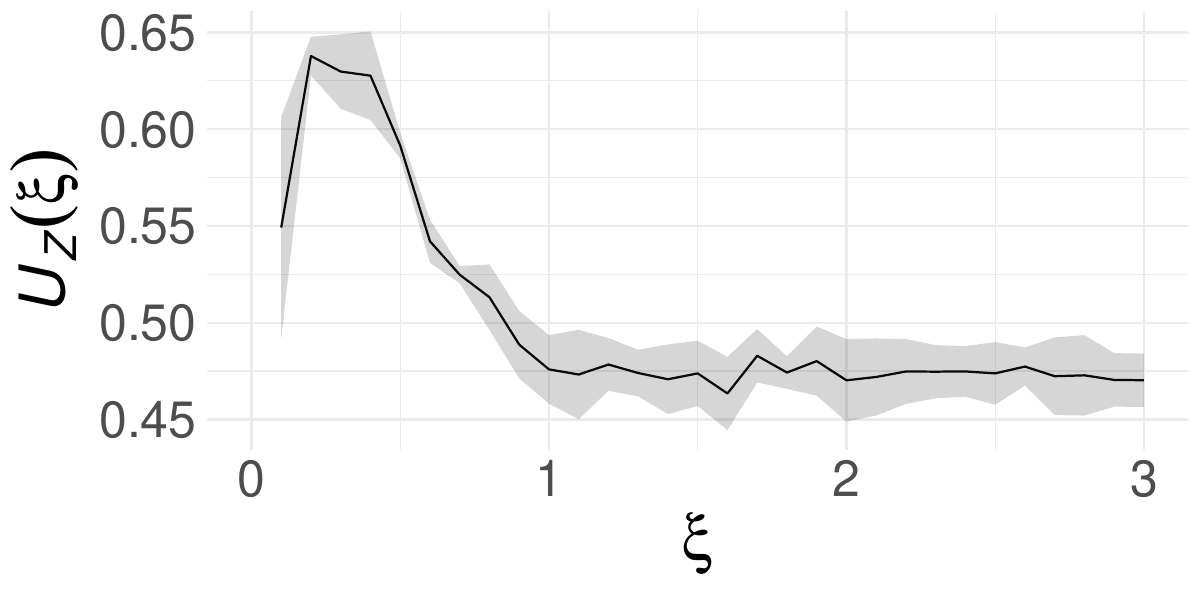}
\qquad
}
 \subfigure[$Z=\beta I(2) S(2)$; optimal at $\xi^{\ast}=2.2, \widehat{U}_{Z,1}(\xi^{\ast})=0.70$]{\label{fig:sirnonlin1}
 \includegraphics[width=0.44\linewidth]{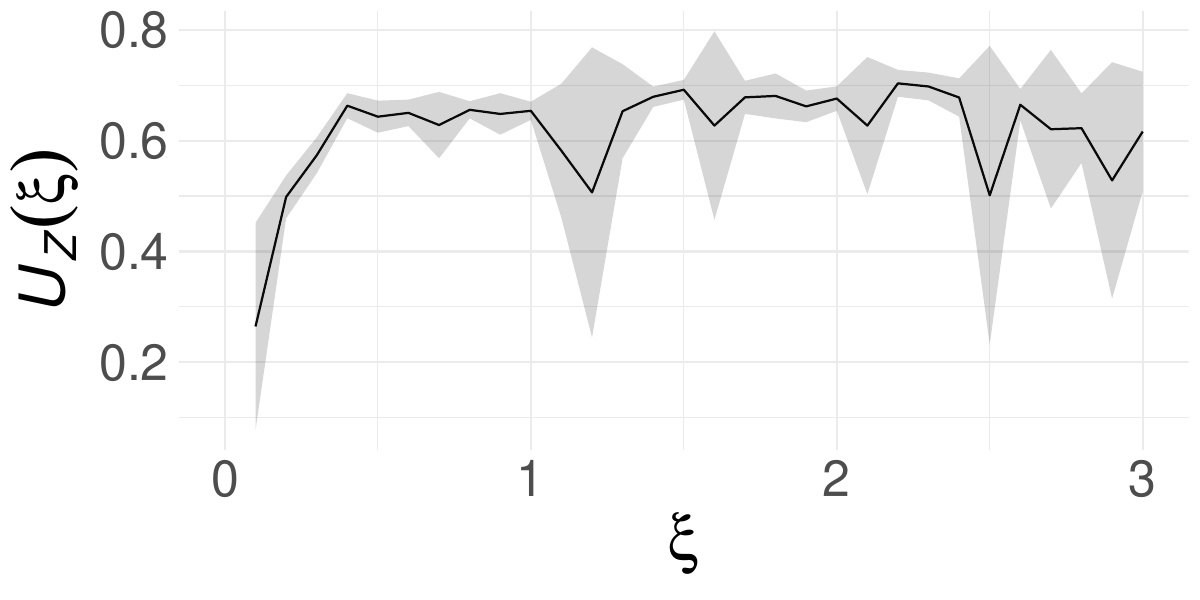}
\qquad
}
 \qquad
 \subfigure[$Z= \beta I(0.1) S(0.1)$; optimal at $\xi^{\ast}=0.4,  \widehat{U}_{Z,1}(\xi^{\ast})=1.15$]{\label{fig:sirnonlin}
 \includegraphics[width=0.44\linewidth]{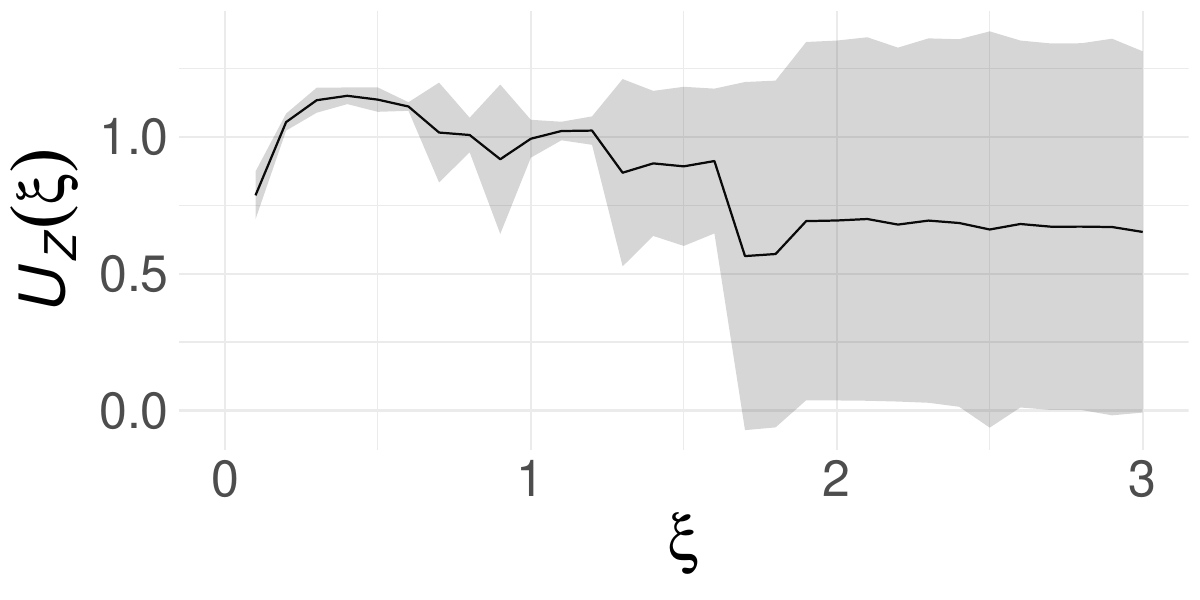}
\qquad}
}
{\caption{$U_Z$ estimated from LF-GO-OED estimator $\widehat{U}_{Z,1}$ for the stochastic SIR example.}}
\end{figure}

We begin with $Z=\Theta$, the non-goal-oriented special case. 
Figure~\ref{fig:sirparaminfnew} shows the mean and one standard deviation interval from five replicates of 
$\widehat{U}_{Z,1}$ plotted across a grid in $\xi\in[0,3]$ (i.e., observation time) with spacing 0.1; each replicate of $\widehat{U}_{Z,1}$ is trained from a new set of 1000 samples. 
Initially when the Infectious population is small, the parameters $\beta$ and $\gamma$ are less informed as state transitions are rare and stochastic. The low $\widehat{U}_{Z,1}$ then starts to increase with time as the number of infected increases. LF-GO-OED finds $\xi^{\ast}=0.4$, achieving a maximum $\widehat{U}_{Z,1}(\xi^{\ast})=2.07.$  When the number of people in the Susceptible and Infectious population start to decrease, $\widehat{U}_{Z,1}$ begins to drop. 

As the nested MC from Section~\ref{ss:nonlinear1D} is not capable of handling implicit likelihood, we implement and compare to the neural network-based NWJ lower bound suggested by \cite{kleinegesse+g21}. With a neural architecture of 2 layers with 20 units per layer, the NWJ lower bound yields an optimal design at 0.46 with an NWJ expected utility estimate of $-0.0055$. Note that $U_Z$, which is the mutual information between $Y$ and $Z$ (see \eqref{eq:goalobjective}), is always non-negative; this NWJ lower bound estimate thus resides in the inadmissible region. 
Upon enriching the neural architecture to 4 layers with 20 units per layer, the optimal design changed to 0.28 with an NWJ expected utility estimate of $-0.00024$. The NWJ-based solution thus appear quite sensitive to the neural architecture, and underscores the challenge of hyperparameter tuning.
For a second comparison, we employed a `brute-force' technique that first uses ABC to sample the parameter posterior, then evaluates the QoI corresponding to each parameter sample, and finally employs KDE to obtain the posterior-predictive---in other words, the same procedure as \citet{zhong+sch22} but replacing MCMC with ABC.
The resulting optimal design is 0.4 with an expected utility estimate of 3.53, much closer to the LF-GO-OED results.

We now turn to the goal-oriented settings, first with a QoI that entails summing the number of recovered at specific time instances of $t=0.3$, $0.4$, and $0.5$: i.e., $Z=R(0.3)+R(0.4)+R(0.5)$. Figure~\ref{fig:sirgoalrecov} shows $\widehat{U}_{Z,1}$ across the design space obtained using the same LF-GO-OED setup, with $\xi^{\ast}=0.2$ and $\widehat{U}_{Z,1}(\xi^{\ast})=0.64$. This contrasts significantly against the optimal design of $\xi^{\ast}=0.4$ from the non-goal-oriented case in Figure~\ref{fig:sirparaminfnew}. 
The disparity emphasizes the importance of adopting the correct design criterion that properly reflects the goal of the experiment.

Lastly, we present results on another QoI: a nonlinear incidence rate represented by $Z = \beta I(t_0) S(t_0)$ at a given time instance $t_0.$ 
This function is motivated by the Monod equation, which generalizes the dependency of a disease's transmission rate on multiple variables like population density, immunity levels, and government interventions \citep{alshammari+k21}. Figures~\ref{fig:sirnonlin1} and \ref{fig:sirnonlin} show $\widehat{U}_{Z,1}$ for $t_0=2$ and $t_0=0.1$, respectively. 
In Figure~\ref{fig:sirnonlin1}, $\widehat{U}_{Z,1}$ is relatively flat for any design greater than $0.4$. At the non-goal-oriented optimal design, i.e., the $\xi^{\ast}=0.4$ from Figure~\ref{fig:sirparaminfnew}, the goal-oriented $\widehat{U}_{Z,1}$ of $0.66$ is very close to the maximum of $0.7$ achieved at the optimal goal-oriented design of $\xi^{\ast}=2.2$. In Figure~\ref{fig:sirnonlin}, the more prominent $\xi^{\ast}$ for this goal appears the same as the non-goal-oriented case. However, there is a heightened variability (the standard deviation interval) for the estimator starting from $\xi=1.3$ onwards. Interestingly, the optimal times for these goals do not align with $t_0$, which one might intuitively expect. We believe this is due to the stochastic and dynamic nature of the system, where observations from a one time might provide more information for a different time. 

\subsection{Stochastic FitzHugh--Nagumo model}\label{sec:fhn}

The FHN model, first introduced by \cite{FitzHugh1961}, is a classic mathematical model in neuroscience for describing the behavior of excitable systems, particularly the generation and propagation of action potentials in neurons. This model has been instrumental in understanding the dynamics of such systems and has far-reaching implications in fields ranging from neurobiology to cardiology.

Following the notations used in \cite{rudi+bl22}, the stochastic FHN model entails a pair of ordinary differential equations (ODEs) describing the evolution of two state variables, the membrane potential $u(t)$ and the recovery variable $v(t)$:
\begin{align}
\frac{{\text{d}u}}{{\text{d}t}} &= \gamma \left(u - \frac{{u^3}}{3} + v + \xi \right),  \\
\frac{{\text{d}v}}{{\text{d}t}} &= -\frac{1}{\gamma}(u - \Theta_0 + \Theta_1v) + \frac{{\text{d}B}}{{\text{d}t}}.
\end{align}
Here $\frac{\text{d}B}{\text{d}t}$ is a stochastic perturbations or variability term that arises from various sources, for instance, current fluctuations, and inherent randomness in biochemical interactions; we model it as a standard Wiener process. 
In the context of OED, the observation variable is $u(t)$ over time, and the design variable $\xi$ is the total membrane current serving as the stimulus applied to the neuron.
Additionally, $\gamma$ is a known constant that governs the level of damping within the system; we set $\gamma=3.0$. The unknown model parameters are $\Theta_0$ and $\Theta_1$. 

In our numerical setup, we set initial conditions $u(0)=0$ and $v(0) = 0$, design space $\xi \in [0,0.8]$ discretized uniformly to 30 points, and priors to be truncated normal distributions: $\Theta_0 \sim \mathcal{TN}(0.4, 0.3^2)$ and $\Theta_1 \sim \mathcal{TN}(0.4, 0.4^2)$ both truncated to within $[0, 1]$.
We observe the time series for $u(t)$ for $t\in [0.1,200]$ at intervals of $0.2$.
Since the time-series is inherently high-dimensional, we employ two summary statistics, spike rate and spike duration \citep{rudi+bl22}, where a spike is defined when the membrane potential $u(t)$ exceeds a threshold of $0.5$ (details on the computation of these statistics see the appendix of \cite{rudi+bl22}).

\begin{figure}[htbp]
    \centering
  \label{fig:fhn}
  {\subfigure[$Z=\Theta$; optimal at $\xi^{\ast}=0.69, \widehat{U}_{Z,1}(\xi^{\ast})=1.94$]{\label{fig:fhnparaminf}%
\includegraphics[width=0.44\linewidth]{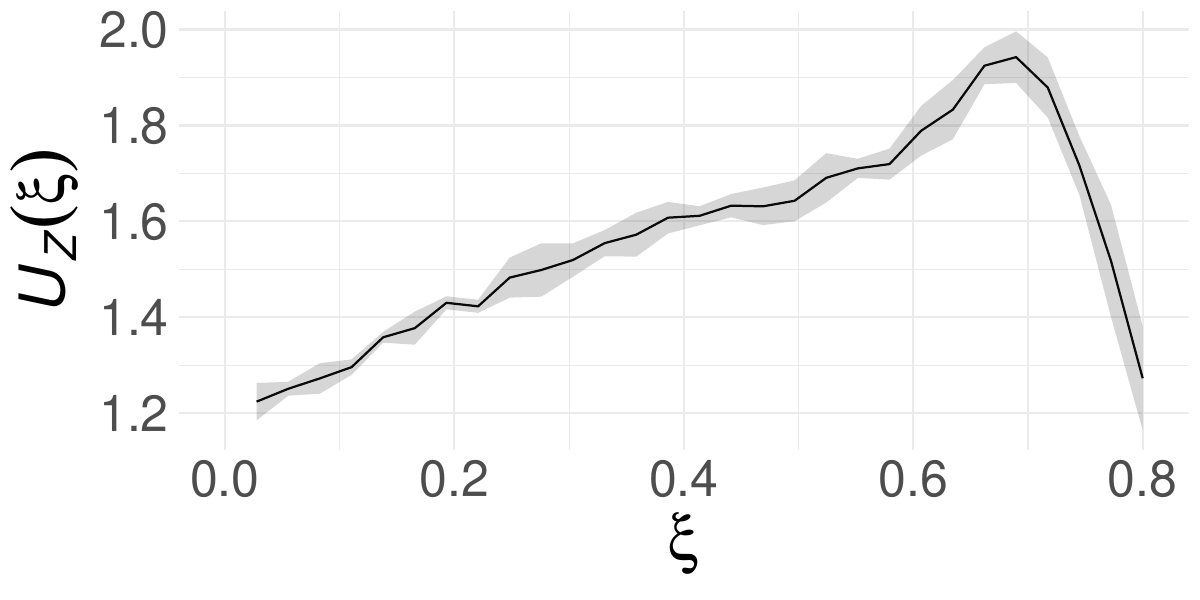}%
    \qquad
    }
    \subfigure[$Z=$ spike rate at a current of 0.2; optimal  at $\xi^{\ast}=0.78, \widehat{U}_{Z,1}(\xi^{\ast})=0.77$]{\label{fig:fhngoal}\includegraphics[width=0.44\linewidth]{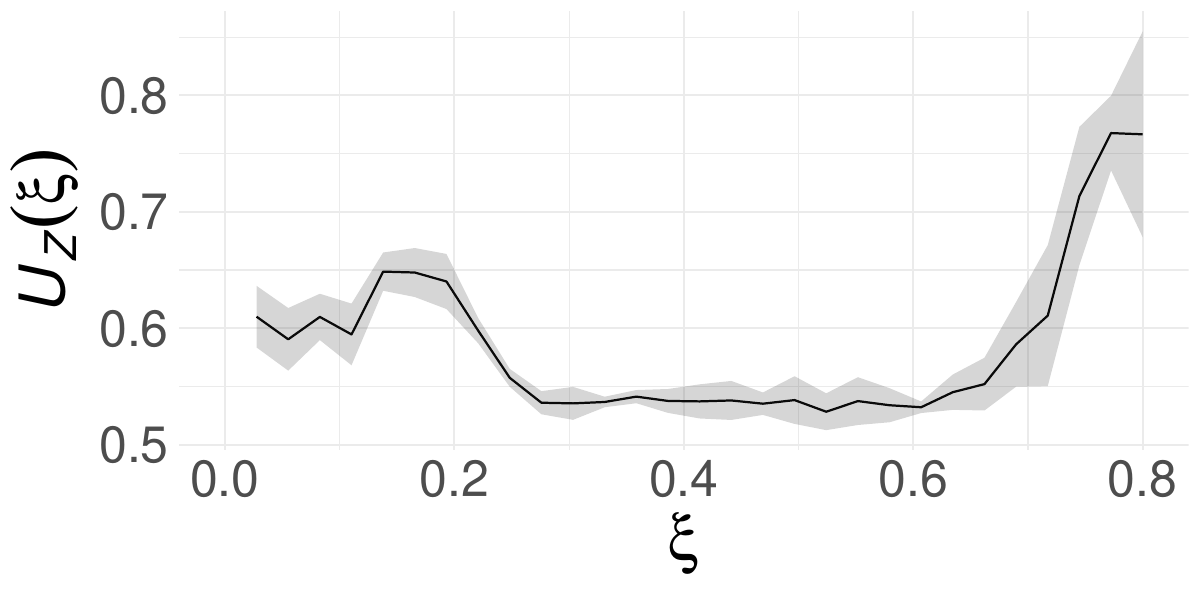}
\qquad}
}
{\caption{$U_Z$ estimates using LF-GO-OED estimator $\widehat{U}_{Z,1}$ for the FHN example.}}
\end{figure}

Again we first show the non-goal-oriented OED results with $Z=\Theta$. 
Figure~\ref{fig:fhnparaminf} shows $\widehat{U}_{Z,1}$ mean and one standard deviation interval obtained from 5 replicates, each trained using 1000 samples. The plot indicates $\xi^{\ast}=0.69$ that achieves $\widehat{U}_{Z,1}(\xi^{\ast})=1.94$. We notice that as the current input increases above around 0.7, the system generates fewer spikes, which we believe accounts for the rapid decrease of $\widehat{U}_{Z,1}$. As a comparison, the NWJ lower bound approach \citep{kleinegesse+g21} using a neural architecture of 2 layers and 20 units per layer identifies an optimal design of 0.4 with an NWJ expected utility estimate of $-0.0055$. Upon enriching  the architecture to 4 layers with 20 units per layer, the optimal design shifts to 0.1 with an NWJ expected utility estimate of $-0.0047$. 
Similar to the SIR example, the optimal design also appears sensitive to the neural architecture choice, and with mutual information estimates that reside in the negative region. 
In a second comparison with the ABC version of \citet{zhong+sch22}, an optimal design is found at 0.717 with an expected utility estimate of 4.737, much closer to the LF-GO-OED results.

We now consider a goal-oriented OED case where the QoI $Z$ is the spike rate at a given current of 0.2. This QoI is of interest because spike rates encode information about neuronal stimulus and their behaviour. Figure~\ref{fig:fhngoal} shows the $\widehat{U}_{Z,1}$ estimates, with the optimal design at $\xi^{\ast}=0.78$ with $\widehat{U}_{Z,1}(\xi^{\ast})=0.77$. Interestingly, the optimal design does not simply coincide the current of 0.2 that will be applied to obtain $Z$; rather, only a small peak of $\widehat{U}_{Z,1}$ exists at a design current of $0.2$. Between design from around $0.3$ to $0.6$, $\widehat{U}_{Z,1}$ for the goal-oriented case remains low and flat, contrasting with the steady increase in the non-goal-oriented case. 
At higher values of design, the goal-oriented case witnesses a sharp increase in $\widehat{U}_{Z,1}$ while the non-goal-oriented cases shows a sharp decline. The non-trivial differences in optimal designs and expected utility curves between the goal-oriented and non-goal-oriented OED further support the importance of of tailoring design criterion to align with the experimental goal.

\section{Conclusion}\label{sec:Conclusion}

We presented LF-GO-OED, a likelihood-free approach to goal-oriented OED for maximizing the EIG on QoIs that depend nonlinearly on the unknown model parameters. 
Central to LF-GO-OED is the training of \emph{density ratio estimators} from samples generated using ABC, which we achieve using the uLSIF algorithm. 
In contrast to previous GO-OED methods, LF-GO-OED can accommodate general nonlinear observation and prediction models, and avoids costly MCMC or density estimation computations.
Notably, LF-GO-OED is suitable for handling intractable likelihoods and does not need any likelihood evaluation or density estimation. 

We validated LF-GO-OED on nonlinear 1D and 2D benchmarks against results from existing and established methods. It was then further deployed to two scientific applications involving intractable likelihoods: the stochastic SIR model from epidemiology and the stochastic FHN model used for neural science. 
The results demonstrated that goal-oriented optimal designs differ significantly both from their non-goal-oriented counterparts and also from our intuition.
By aligning the design criterion towards the experimental goals, resulting designs can be more relevant to the scientific investigation of the complex system in the experiment.

One limitation of LF-GO-OED is the computationally taxing process of training the density ratio estimator for each candidate design and data realization. An avenue of future research thus entails developing accurate amortized versions of the density ratio estimator. 
Additionally, ABC for OED is effective mostly in low-dimensional design spaces and requires designing or selection of the summary statistics, distance metric, and threshold. 
Efficient, more scalable alternatives to ABC are therefore also of interest. 

\medskip
\section{Acknowledgements}

The authors express their gratitude to Johann Rudi for his valuable insights on the FHN example.
\medskip
\section{Funding}

This material is based upon work supported by the U.S. Department of Energy, Office of Science, Office of Advanced Scientific Computing and Sandia National Laboratories Advanced Science and Technology  Laboratory Directed Research and Development program. Sandia National Laboratories is a multimission laboratory managed and operated by National Technology and Engineering Solutions of Sandia, LLC., a wholly owned subsidiary of Honeywell International, Inc., for the U.S. Department of Energy's National Nuclear Security Administration under contract DE-NA0003525. This paper describes objective technical results and analysis. Any subjective views or opinions that might be expressed in the paper do not necessarily represent the views of the U.S. Department of Energy or the United States Government.

This research used resources of the National Energy Research Scientific Computing Center (NERSC), a U.S. Department of Energy Office of Science User Facility located at Lawrence Berkeley National Laboratory, operated under Contract No. DE-AC02-05CH11231 using NERSC award DOE-ERCAPm3876.

This research was supported in part through computational resources and services provided by Advanced Research Computing at the University of Michigan, Ann Arbor.

The authors report there are no competing interests to declare.

\section{Code}
Source code can be found at \url{https://github.com/atlantac96/LF-GO-OED.git}.

%\bigskip

\appendix

\section{Appendix}

\subsection{ABC Threshold}

The threshold used in ABC, $\epsilon$, is selected from a sequence of values ranging from 0.1 to 0.5. We perform leave-one-out cross validation to identify the setting associated with the lowest prediction error 
using the \texttt{cv4abc} package in R \citep{abc}. Through computations across all our examples, we find that generally a small threshold results in insufficient variation in the summary statistics. It is important to note that this threshold selection process can be time-consuming for our neural net adjustment ABC fitting. 

Table~\ref{tab:thresholdabc} shows the ABC threshold versus cross validation prediction error along with their computational times. All computations were carried out using 36 cores of dual 3.0 GHz Intel Xeon Gold 6154 processors. For the SIR example, we observe the cross validation error to increase with threshold value; we ultimately choose a threshold of 0.1. For the FHN example, the error decreases with threshold value; we ultimately choose a threshold of 0.5. The error values in the FHN case also significantly exceed those in the SIR case due to the utilization of summary statistics for FHN.

\begin{table}[htb]
\centering
 \caption{Median prediction error of 1000 samples based on a leave-one-out cross validation. (Left) SIR case, took 19.18 mins, and (right) FHN case, took 22.11 mins.}
 {%
   \subtable{%
     \label{tab:ab}%
     \begin{tabular}{|c|c|c|}
        \hline 
Tolerance, $\epsilon$ & $\beta$ & $\gamma$ \\
\hline 

0.1 & 0.0645 & 0.0256 \\
0.2 & 0.0673 & 0.0273 \\
0.3 & 0.0742 & 0.0331 \\
0.4 & 0.0892 & 0.0477 \\
0.5 & 0.1000 & 0.0597 \\
\hline
        \end{tabular}
   }\qquad
   \subtable{%
     \label{tab:cd}%
     \begin{tabular}{|c|c|c|}
            \hline
       Tolerance, $\epsilon$ & $\Theta_0$ & $\Theta_1$ \\
        \hline

0.1 & 0.3033 & 0.5953 \\
0.2 & 0.2957 & 0.5759 \\
0.3 & 0.2929 & 0.5737 \\
0.4 & 0.2888 & 0.5732 \\
0.5 & 0.2845 & 0.5593 \\
\hline
        \end{tabular}
   }
 }
  \label{tab:thresholdabc}
\end{table}

\subsection{Computational Complexity}

Here we investigate how LF-GO-OED's computational time scales with the design space dimension
through an extended version of the 2D benchmark in Section~\ref{ss:2D}. 
In this setup, the observation data, parameter, and design spaces are all $N$-dimensional, with
\begin{align}
Y_i = \Theta_i^3\xi_i^2+ \sum_{j \in \{1,\ldots,N\}, j \neq i}\Theta_j \exp{-|0.2-\xi_j|} + \mathcal{E}_i, \quad i=1,\ldots,N,
\end{align} 
where $\mathcal{E}_i \sim \mathcal{N}(0,10^{-4})$ and prior $\Theta_i\sim \mathcal{U}([0,1])$ are all independent, and $\xi_i\in[0,1]$.
The QoI is set to be the Rosenbrock function: 
\begin{align}
Z=\sum_{i=1}^N(1-\Theta_i)^2+\sum_{j \in \{1,\ldots,N\}, j\neq i}5(\Theta_j-\Theta_i^2)^2.
\end{align}
For our analysis, we consider 500 samples for training $\widehat{U}_{Z,1}$. 
Since the design dimension is also increasing, gridding the expected utility in the design space would no longer be feasible. Instead, we employ Bayesian optimization~\citep{shahriari2015taking,frazier2018tutorial} to solve \eqref{optimal design}. 

\begin{figure}[h!]
    \centering
    \includegraphics{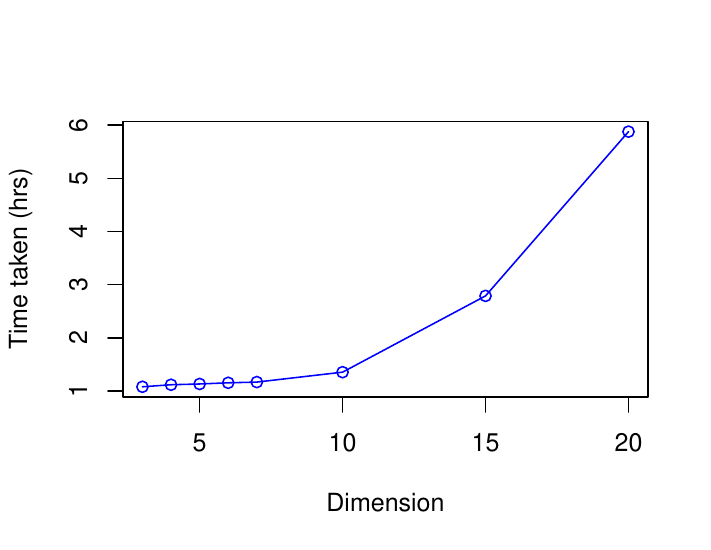}
    \caption{LF-GO-OED computational time versus changes in observation data, parameter, and design space dimension.}
    \label{fig:highdesign}
\end{figure}

Table \ref{tab:dimensionvstime} presents the computational time versus the design dimension, while Figure~\ref{fig:highdesign} illustrates this graphically; all times are reported when using 36 cores of dual 3.0 GHz Intel Xeon Gold 6154 processors. Notably, uLSIF showed a noted decline in accuracy beyond 10 dimensions. For the 20-dimensional case, the reported time was acquired after 7 epochs of Bayesian optimization, and full convergence might not have been achieved.

\begin{table}[htb]
    \centering
    \begin{tabular}{|c|c|}
    \hline
       Dimension $N$ & Computational Time (hrs) \\
       \hline
         
         3 & 1.0772 \\
4 & 1.1157 \\
5 & 1.1300 \\
6 & 1.1523 \\
7 & 1.1660 \\
10 & 1.3524 \\
15 & 2.7875 \\
20 & 5.8779 \\
         \hline
    \end{tabular}
    \caption{LF-GO-OED computational time versus changes in observation data, parameter, and design space dimension.}
    \label{tab:dimensionvstime}
\end{table} 

We also investigate the time required to evaluate the expected utility at one design point using LF-GO-OED with respect to the number of MC samples, via the 1D benchmark in Section~\ref{ss:nonlinear1D}. Table~\ref{tab:samplesvstime} shows these results; the time increases approximately 44 fold when the number of samples increased from 1000 to 5000, and approximately 412 fold when the number of samples increased from 1000 to 10000; computational time is roughly exponential with the number of samples.
In contrast, for 1000 samples in 1D, where the likelihood is explicit, the nested MC estimator at one design point takes about 5.423 seconds---this highlights the substantial increase in difficulty for implicit likelihood OED problems.

 \begin{table}[h!]
    \centering
    \begin{tabular}{|c|c|}
    \hline
       Number of Samples  & Computational Time (hrs) \\
       \hline
        
        1000 & 0.0239\\
        5000 & 1.0561\\
        10000 & 9.8556\\
         \hline
    \end{tabular}
    \caption{LF-GO-OED computational time versus number of samples used for evaluation at one design point.}
    \label{tab:samplesvstime}
\end{table}

%\bibliographystyle{chicago}

%\bibliography{references}

\end{document}